\documentclass[prd,aps,floats,showpacs,twocolumn,twoside,preprintnumbers,
superscriptaddress,floatfix]{revtex4}
\usepackage{bm}
\usepackage[T1]{fontenc}
\usepackage[latin1]{inputenc}
\usepackage{amsmath}
\usepackage{amssymb}
\usepackage{mathrsfs}
\usepackage{epsfig}

\newcommand{\pslash}{{p\!\!\!/}}
\newcommand{\dslash}{{\partial\!\!\!/}}

\newcommand\expect[1]{\langle\!\langle#1\rangle\!\rangle}
\newcommand\eqn[1]{(\ref{#1})}
\newcommand\eqns[2]{(\ref{#1}-\ref{#2})}

\begin{document}



\title{The phase diagram of three-flavor quark matter under compact star 
constraints}

\author{D. Blaschke}
\email{Blaschke@theory.gsi.de}
\affiliation{Gesellschaft f\"ur Schwerionenforschung mbH (GSI),
D-64291 Darmstadt, Germany, and\\
Bogoliubov Laboratory for Theoretical Physics, JINR Dubna, 141980 Dubna, 
Russia}
\author{S. Fredriksson}
\email{Sverker.Fredriksson@ltu.se}
\affiliation{Department of Physics, Lule{\aa} University of Technology,
SE-97187 Lule\aa , Sweden}
\author{H. Grigorian}
\email{Hovik.Grigorian@uni-rostock.de}
\affiliation{Institut f\"ur Physik, Universit\"at Rostock, 
D-18051 Rostock, Germany, and\\
Department of Physics, Yerevan State University,375025 Yerevan, Armenia}
\author{A.M. \"Ozta\c{s}}
\email{oztas@hacettepe.edu.tr}
\affiliation{Department of Physics, Hacettepe University,
TR-06532 Ankara, Turkey}
\author{F. Sandin}
\email{Fredrik.Sandin@ltu.se}
\affiliation{Department of Physics, Lule{\aa} University of Technology,
SE-97187 Lule\aa , Sweden}


\begin{abstract}
The phase diagram of three-flavor quark matter under compact star
constraints is investigated within a Nambu--Jona-Lasinio model.
Local color and electric charge neutrality is imposed for
$\beta$-equilibrated superconducting quark matter. The constituent
quark masses and the diquark condensates are determined selfconsistently
in the plane of temperature and quark chemical potential. 
Both strong and intermediate diquark coupling strengths are considered. 
We show that in both cases, gapless superconducting phases
do not occur at temperatures relevant for compact star evolution, {i.e.},
below $T \sim 50$ MeV. The stability and structure of isothermal quark
star configurations are evaluated. 
For intermediate coupling, quark stars are composed of a mixed phase of
normal (NQ) and two-flavor superconducting (2SC) quark matter up to a 
maximum mass of $1.21~M_\odot$.
At higher central densities, a phase transition to the three-flavor 
color flavor locked (CFL) phase occurs and the configurations become unstable.
For the strong diquark coupling we find stable stars in the 2SC phase, with 
masses up to $1.326~M_\odot$. A second family of more compact
configurations (twins) with a CFL quark matter core and a 2SC shell is also 
found to be stable. The twins have masses in the range $1.301 ... 1.326$ 
$M_\odot$. We consider also hot isothermal configurations at temperature 
$T=40$ MeV. When the hot maximum mass configuration cools down, due to 
emission of photons and neutrinos, a mass defect of $0.1~M_\odot$ occurs 
and two final state configurations are possible.
\end{abstract}


\maketitle


\section{Introduction}

Theoretical investigations of the QCD phase diagram at high densities
have recently gained momentum due to results of non-perturbative
low-energy QCD models~\cite{rapp98,alford99,Blaschke:1998md} of
color superconductivity in quark matter~\cite{barrois77,bailin84}.
These models predict that the diquark pairing condensates are of the
order of 100 MeV and a remarkably rich phase structure has been
identified~\cite{Rajagopal:2000wf,Alford:2001dt,Buballa:2003qv,Schmitt:2004et}.
The main motivation for studying the low-temperature domain of the 
QCD phase diagram is its possible relevance for the physics of 
compact stars~\cite{NSI,CS,SQCDM}. Observable effects of color superconducting
phases in compact stars is expected, {e.g.}, in the
cooling behaviour~\cite{Horvath:1991ms,Blaschke:1999qx,Page:2000wt,Blaschke:2000dy,Grigorian:2004jq},
magnetic field evolution~\cite{Blaschke:1999fy,Alford:1999pb,Iida:2002ev,Sedrakian:2002ff},
and in burst-type phenomena~\cite{Hong:2001gt,Ouyed:2001cg,Aguilera:2002dh,Ouyed:2005tm}.

The most prominent color superconducting phases with large diquark 
pairing gaps are the two-flavor scalar diquark condensate (2SC) and 
the color-flavor locking (CFL) condensate. The latter requires 
approximate SU(3) flavor symmetry and occurs therefore only at rather 
large quark chemical potentials, $\mu_q>430-500$ MeV, of the order of the 
dynamically generated strange quark mass $M_s$, whereas the 2SC phase can 
appear already at the chiral restoration transition for  $\mu_q>330-350$ MeV
\cite{Neumann:2002jm,Oertel:2002pj,Gocke:2001ri}. 
Note that the quark chemical potential in the center of a typical compact 
star is expected to not exceed a value of $\sim 500$ MeV so
the volume fraction of a strange quark matter phase will be insufficient 
to entail observable consequences.
However, when the strange quark mass is considered not dynamically, but 
as a free parameter independent of the thermodynamical conditions, it has been 
shown that for not too large $M_s$ the CFL phase dominates over the 2SC phase
\cite{Alford:2002kj,Steiner:2002gx}. 
Studies of the QCD phase diagram with fixed strange quark mass have recently 
been extended to the discussion of gapless CFL (gCFL) phases 
\cite{Alford:2004akr,Alford:2004akr2,Ruster:2004}.
The gapless phases occur when the asymmetry between Fermi levels of different 
flavors is large enough to allow for zero energy excitations while a finite
pairing gap exists. They have been found first for the 2SC phase (g2SC) 
within a dynamical chiral quark model \cite{Shovkovy:2003uu,Huang:2003}.

Any scenario for compact star evolution 
that is based on the occurence of quark matter relies 
on the assumptions about the properties of this phase. 
It is therefore of prior importance to obtain a phase diagram of three-flavor 
quark matter  under compact star constraints with selfconsistently
determined dynamical quark masses. 
In the present paper we will employ the Nambu--Jona-Lasinio (NJL) model
to delineate the different quark matter phases in the plane of temperature
and chemical potential. We also address the question whether CFL quark matter 
and gapless phases  are likely to play a role in compact star interiors. 


\section{Model}
In this paper, we consider an NJL model with quark-antiquark interactions in 
the color singlet scalar/pseudoscalar channel, and quark-quark interactions in 
the scalar color antitriplet channel.
We neglect the less attractive interaction channels, {e.g.}, the 
isospin-singlet channel, which could allow for weak spin-1 condensates. 
Such condensates allow for gapless excitations at low temperatures and could 
be important for the cooling behaviour of compact stars. 
However, the coupling strengths in these channels are poorly known and we 
therefore neglect them here.
The Lagrangian density is is given by
\begin{eqnarray}
\label{lagrangian}
        {\cal L} &=& \bar{q}_{i\alpha}(i\dslash\delta_{ij}\delta_{\alpha\beta} 
- M^0_{ij}\delta_{\alpha\beta}
        + \mu_{ij,\alpha\beta}\gamma^0)q_{j\beta}  \nonumber \\
        &+& G_S \sum^8_{a=0}\left[(\bar{q}\tau^a_f q)^2 + 
 (\bar{q}i\gamma_5\tau^a_f q)^2\right] \nonumber \\
        &+&\,G_{D} \sum_{k, \gamma} \left[(\bar{q}_{i\alpha}\epsilon_{ijk}
\epsilon_{\alpha\beta\gamma}q^C_{j\beta})
        (\bar{q}^C_{i'\alpha'}\epsilon_{i'j'k}\epsilon_{\alpha'\beta'\gamma}
q_{j'\beta'}) \right. \nonumber \\
        &+& \left. (\bar{q}_{i\alpha}i\gamma_5\epsilon_{ijk}
\epsilon_{\alpha\beta\gamma}q^C_{j\beta})
        (\bar{q}^C_{i'\alpha'}i\gamma_5\epsilon_{i'j'k}
\epsilon_{\alpha'\beta'\gamma}q_{j'\beta'}) \right],
\end{eqnarray}
where $M^0_{ij}=$~diag$(m^0_u, m^0_d, m^0_s)$ is the current quark mass
matrix in flavor space and $\mu_{ij,\alpha\beta}$ is the chemical
potential matrix in color and flavor space. Due to strong and weak
interactions, the various chemical potentials are not independent. In the
superconducting phases a $U(1)$ gauge symmetry remains unbroken~
\cite{Alford:1999arw}, and the associated charge is a linear combination of 
the electric charge, $Q$, and two orthogonal generators of the unbroken 
$SU(2)_c$ symmetry. Hence, there are in total four independent chemical 
potentials
\begin{equation}
\label{chempot}
\mu_{ij,\alpha\beta} = (\mu\delta_{ij} + Q\mu_Q)\delta_{\alpha\beta} + 
(T_3\mu_3 + T_8\mu_8)\delta_{ij},
\end{equation}
where $Q={\rm diag}(2/3,-1/3,-1/3)$ is the electric charge in flavor space,
and $T_3={\rm diag}(1,-1,0)$ and $T_8={\rm diag}(1/\sqrt{3},1/\sqrt{3},
-2/\sqrt{3})$
are the generators in color space. The quark number chemical
potential, $\mu$, is related to the baryon chemical potential by 
$\mu = \mu_B/3$.
The quark fields in color, flavor and Dirac spaces are denoted by
$q_{i\alpha}$ and $\bar{q}_{i\alpha}=q^\dagger_{i\alpha}\gamma^0$.
$\tau^a_f$ are Gell-Mann matrices acting in
flavor space. Charge conjugated quark fields are denoted by 
$q^C = C \bar{q}^T$ and $\bar{q}^C=q^T C$, where $C=i\gamma^2\gamma^0$ 
is the Dirac charge conjugation matrix.
The indices $\alpha$, $\beta$ and $\gamma$ represent colors 
($r=1$, $g=2$ and $b=3$),
while $i$, $j$ and $k$ represent flavors ($u=1$, $d=2$ and $s=3$). 
$G_S$ and $G_D$ are dimensionful coupling constants that must be determined 
by experiments.

Typically, three-flavor NJL models use a 't Hooft determinant interaction 
that induces a U$_A$(1) symmetry breaking in the pseudoscalar isoscalar 
meson sector which can be adjusted such that the $\eta$-$\eta'$ mass 
difference is described.
This realization of the   U$_A$(1) breaking leads to the important consequence
that the quark condensates of different flavor sectors get coupled. 
The dynamically generated strange quark mass contains a contribution from the
chiral condensates of the light flavors.
There is, however, another possible realization of the U$_A$(1) symmetry 
breaking that does not arise on the mean field level, but only for the mesonic 
fluctuations in the pseudoscalar isoscalar channel. This is due to the
coupling to the nonperturbative gluon sector via the the triangle anomaly,
see e.g. \cite{Blaschke:1996dp,vonSmekal:1997dq,Nielsen:1992va}.
This realization of the $\eta$-$\eta'$ mass difference gives no contribution 
to the quark thermodynamics at the mean field level, which we 
will follow in this paper.
Up to now it is not known, which of the two U$_A$(1) breaking mechanisms is the
dominant one in nature.
In the present exploratory study of the mean field thermodynamics
of three-flavor quark matter, we will take the point of view that the 't Hooft 
term might be subdominant and can be disregarded. 
One possible way to disentangle both mechanisms is due to their different 
response to chiral symmetry restoration at finite temperatures and densities.
While in heavy-ion collisions only the finite temperature aspect can be 
systematically studied \cite{AAR}, the state of matter in neutron star 
interiors may be suitable to probe the U$_A$(1) symmetry restoration and 
its possible implications for the quark matter phase diagram at high 
densities and low temperatures. 
A comparison of the results presented in this work with 
the alternative treatment of the phase diagram of three-flavor quark 
matter including the 't Hooft determinant term, see \cite{RWBSR}, may 
therefore be very instructive.

The mean-field Lagrangian is
\begin{widetext}
\begin{eqnarray}
        {\cal L}^{MF} &=& \bar{q}_{i\alpha}\left[i\dslash\delta_{ij}\delta_{\alpha\beta}
        - (M^0_{ij} - 4 G_S \expect{\bar{q}_{i\alpha}q_{j\beta}}\delta_{ij})\delta_{\alpha\beta}
        + \mu_{ij,\alpha\beta}\gamma^0 \right]q_{j\beta} \nonumber \\
        &-& 2 G_S \sum_i \expect{\bar{q}_i q_i}^2
        - \sum_{k, \gamma} \frac{|\Delta_{k\gamma}|^2}{4 G_D} + \bar{q}_{i\alpha}\frac{\widetilde{\Delta}_{k\gamma}}{2}
        q^C_{j\beta} + \bar{q}^C_{i\alpha}\frac{\widetilde{\Delta}^\dagger_{k\gamma}}{2} q_{j\beta},
        \label{mflagrangian} \\
        \widetilde{\Delta}_{k\gamma} &=& 2 G_D i \gamma_5\epsilon_{\alpha\beta\gamma}\epsilon_{ijk}
        \expect{\bar{q}_{i'\alpha'}i\gamma_5\epsilon_{\alpha'\beta'\gamma}\epsilon_{i'j'k} q^C_{j'\beta'}}
        = i\gamma_5\epsilon_{\alpha\beta\gamma}\epsilon_{ijk} \Delta_{k\gamma}.
        \label{digap}
\end{eqnarray}
\end{widetext}
We define the chiral gaps
\begin{equation}
\label{chiralcond}
        \phi_i = -4G_S\expect{\bar{q_i}q_i},
\end{equation}
and the diquark gaps
\begin{equation}
\label{diqcond}
        \Delta_{k\gamma} = 2G_D\expect{\bar{q}_{i\alpha}i\gamma_5\epsilon_{\alpha\beta\gamma}\epsilon_{ijk} q^C_{j\beta}}.
\end{equation}

The chiral condensates contribute to the dynamical mass of the quarks, the constituent quark
mass matrix in flavor space is $M = \text{diag}(m^0_u+\phi_u, m^0_d+\phi_d, m^0_s+\phi_s)$,
where $m^0_i$ are the current quark masses.
For finite current quark masses the $U(3)_L \times U(3)_R$ symmetry of the Lagrangian is
spontaneously broken and only approximately restored at high densities.

The diquark gaps, $\Delta_{k\gamma}$, are antisymmetric in flavor and color, 
{ e.g.}, the condensate corresponding to $\Delta_{ur}$ is created by green 
and blue down and strange quarks. Due to
this property, the diquark gaps can be denoted with the flavor indices of the
interacting quarks
\begin{equation}
        \Delta_{ur}=\Delta_{ds}, \quad \Delta_{dg}=\Delta_{us}, 
\quad \Delta_{sb}=\Delta_{ud}.
\label{digapidents}
\end{equation}

After reformulating the mean-field lagrangian in 8-component Nambu-Gorkov
spinors~\cite{Gorkov:1959,Nambu:1960} and performing the functional integrals
over Grassman variables~\cite{Kapusta:1989} we obtain the thermodynamic 
potential
\begin{eqnarray}
        \label{matsubpotential}
        \Omega(T,\mu) &=& \frac{\phi^2_u+\phi^2_d+\phi^2_s}{8 G_S}
        +\frac{|\Delta_{ud}|^2+|\Delta_{us}|^2+|\Delta_{ds}|^2}{4 G_D} 
\nonumber \\
        &-&T \sum_n \int\frac{d^3p}{(2\pi)^3}\frac{1}{2}{\rm Tr}
\ln\left(\frac{1}{T}S^{-1}(i\omega_n,\vec{p})\right) 
\nonumber \\
        &+& \Omega_e - \Omega_0.
\end{eqnarray}
Here $S^{-1}(p)$ is the inverse propagator of the quark fields at four momentum
$p=(i\omega_n,\vec{p})$,
\begin{equation}
       S^{-1}(i\omega_n,\vec{p}) = \left[
       \begin{array}{cc}
               \pslash-M+\mu\gamma^0&\widetilde{\Delta}_{k\gamma} \\
               \widetilde{\Delta}_{k\gamma}^\dagger&\pslash-M-\mu\gamma^0
       \end{array}
       \right],
\label{invprop}
\end{equation}
and $\omega_n = (2n+1)\pi T$ are the Matsubara frequencies for fermions. 
The thermodynamic potential of ultrarelativistic electrons,
\begin{equation}
\label{omegae}
        \Omega_e = -\frac{1}{12\pi^2}\mu^4_Q -\frac{1}{6}\mu^2_QT^2 - 
                \frac{7}{180}\pi^2T^4,
\end{equation}
has been added to the potential, and the vacuum contribution,
\begin{eqnarray}
\label{omega0}
        \Omega_0 = \Omega(0,0) &=& 
                \frac{\phi^2_{0u}+\phi^2_{0d}+\phi^2_{0s}}{8 G_S} 
\nonumber \\
        && - 2 N_c\sum_i\int\frac{d^3p}{(2\pi)^3}\sqrt{M^2_i + p^2},
\end{eqnarray}
has been subtracted in order to get zero pressure in vacuum. 
Using the identity $\text{Tr}(\text{ln}(D))=\text{ln}(\text{det}(D))$
and evaluating the determinant (see Appendix A), we obtain   
\begin{equation}
        \text{ln}\;\text{det}\left(\frac{1}{T}S^{-1}(i\omega_n,\vec{p})\right)=
        2\sum_{a=1}^{18}\text{ln}\left(\frac{\omega^2_n + \lambda_a(\vec{p})^2}{T^2}\right).
\end{equation}
The quasiparticle dispersion relations, $\lambda_a(\vec{p})$, are the eigenvalues
of the Hermitian matrix,
\begin{equation}
        {\cal M} = \left[
        \setlength\arraycolsep{-0.01cm}
        \begin{array}{cc}
                -\gamma^0\vec{\gamma}\cdot\vec{p}-\gamma^0 M+\mu & \gamma^0\widetilde{\Delta}_{k\gamma}C \\
                \gamma^0 C\widetilde{\Delta}_{k\gamma}^\dagger & -\gamma^0 \vec{\gamma}^T\cdot\vec{p}+\gamma^0 M-\mu
        \end{array}
        \right],
\label{eigmatrix}
\end{equation}
in color, flavor, and Nambu-Gorkov space. This result is in agreement with~
\cite{Steiner:2002gx,RWBSR}. 
Finally, the Matsubara sum can be evaluated on closed form~\cite{Kapusta:1989},
\begin{equation}
\label{matsubarasum}
    T\sum_n\ln\left(\frac{\omega^2_n + \lambda^2_a}{T^2}\right) 
=\lambda_a  + 2 T\ln(1+e^{-\lambda_a/T}),
\end{equation}
leading to an expression for the thermodynamic potential on the form 
\begin{eqnarray}
        \Omega(T,\mu) &=& \frac{\phi^2_u+\phi^2_d+\phi^2_s}{8 G_S}
        +\frac{|\Delta_{ud}|^2+|\Delta_{us}|^2+|\Delta_{ds}|^2}{4 G_D} 
\nonumber \\
     &-&\int\frac{d^3p}{(2\pi)^3}\sum_{a=1}^{18}
         \left(\lambda_a+2T\ln\left(1+e^{-\lambda_a/T}\right)\right)
\nonumber \\
        &+& \Omega_e - \Omega_0.
\label{potential}
\end{eqnarray}
It should be noted that (14) is an even function
of $\lambda_a$, so the signs of the quasiparticle
dispersion relations are arbitrary. In this paper,
we assume that there are no trapped neutrinos. This
approximation is valid for quark matter in neutron
stars, after the short period of deleptonization
is over.

Equations~\eqn{omegae},~\eqn{omega0},~\eqn{eigmatrix}, and~\eqn{potential} 
form a
consistent thermodynamic model of superconducting quark matter. The independent
variables are $\mu$ and $T$. The gaps, $\phi_i$, and $\Delta_{ij}$, are
variational order parameters that should be determined by minimization of the
grand canonical thermodynamical potential, $\Omega$. Also, quark matter should
be locally color and electric charge neutral, so at the physical minima of the
thermodynamic potential the corresponding number densities should be zero
\begin{eqnarray}
        n_Q &=& -\frac{\partial\Omega}{\partial\mu_Q} = 0, \\
        n_8 &=& -\frac{\partial\Omega}{\partial\mu_3} = 0, \\
        n_3 &=& -\frac{\partial\Omega}{\partial\mu_8} = 0.
\end{eqnarray}
The pressure, $P$, is related to the thermodynamic potential by $P = -\Omega$
at the global minima of $\Omega$.
The quark density, entropy and energy density are then obtained 
as derivatives of the thermodynamical potential with respect to $\mu$, $T$ and
$1/T$, respectively.


\section{Results}
The numerical solutions to be reported in this Section are obtained with 
the following set of model parameters, taken from Table 5.2 of Ref. 
\cite{Buballa:2003qv} for vanishing 't Hooft interaction,
\begin{eqnarray}
m^0_{u,d}&=& 5.5~{\mathrm MeV}~,\\
m^0_s &=& 112.0~ {\mathrm MeV}~,\\
G_S \Lambda^2 &=& 2.319~,\\
\Lambda^2 &=& 602.3~ {\mathrm MeV}~.
\end{eqnarray}
With these parameters, the following low-energy QCD observables can be 
reproduced: $m_\pi=135$ MeV, $m_K=497.7$ MeV, $f_\pi=92.4$ MeV.
The value of the diquark coupling strength $G_D=\eta G_S$ is considered as 
a free parameter of the model. Here we present results for $\eta=0.75$ 
(intermediate coupling) and $\eta=1.0$ (strong coupling). 

\begin{figure}[thb]
\includegraphics[width=0.45\textwidth,angle=-90]{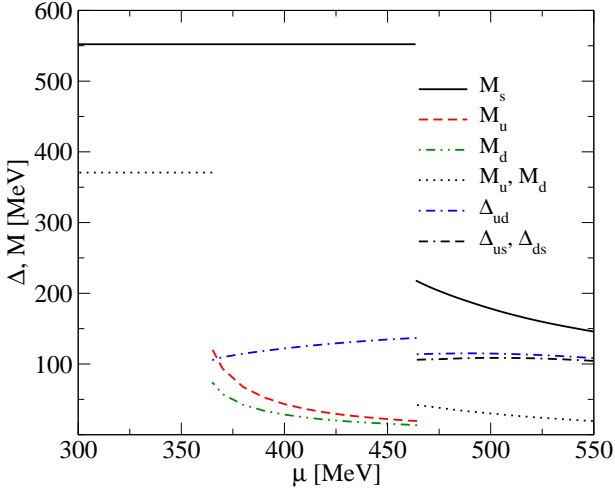}
\caption{Gaps and dynamical quark masses as a function of $\mu$ at T=0 for
intermediate diquark coupling, $\eta=0.75$.
\label{gaps075}
}
\end{figure}
\begin{figure}[thb]
\label{gaps1}
\vspace{-5mm}  
\includegraphics[width=0.45\textwidth,angle=-90]{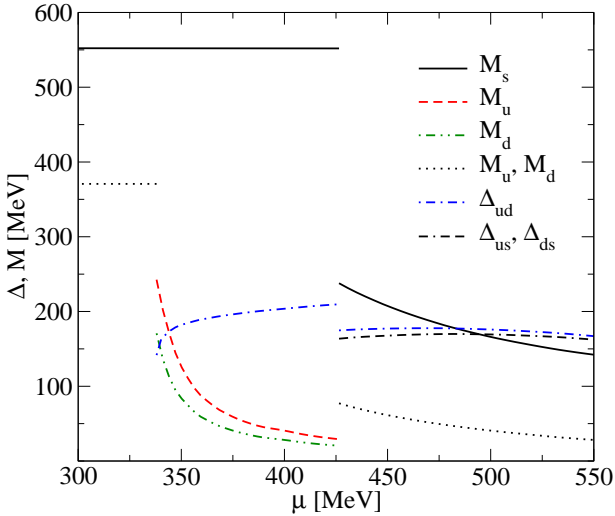}
\caption{Gaps and dynamical quark masses as a function of $\mu$ at T=0 for
strong diquark coupling, $\eta=1$.}
\end{figure}

\subsection{Quark masses and pairing gaps at zero temperature}
The dynamically 
generated quark masses and the diquark pairing gaps are determined
selfconsistently at the absolute minima of the thermodynamic potential,
in the plane of temperature and quark chemical potential.
This is done for both the strong  and the intermediate diquark coupling 
strength.
In Figs. 1 and 2 we show the dependence of 
masses and gaps on the quark chemical potential at $T=0$ for $\eta=0.75$ and
$\eta=1.0$, resp. 
\begin{figure}[hbt]
\vspace{-5mm}  
\includegraphics[width=0.45\textwidth,angle=-90]{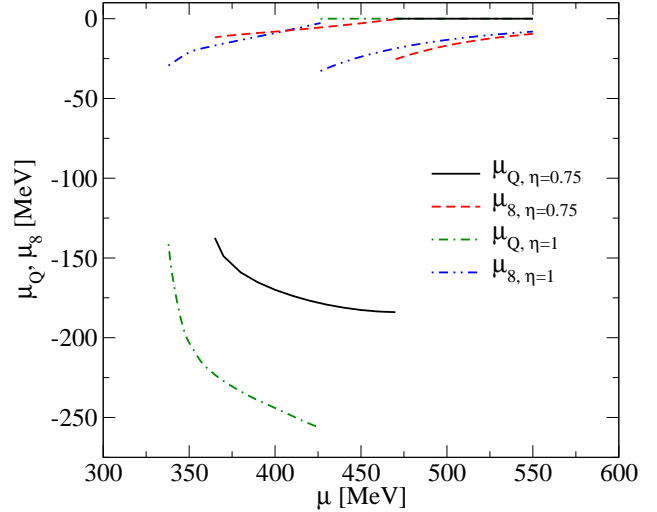}
\caption{Chemical potentials $\mu_Q$ and $\mu_8$ at T=0 for
both values of the diquark coupling $\eta=0.75$ and $\eta=1$.
All phases considered in this work have zero
$n_3$ color charge for $\mu_3=0$, hence $\mu_3$
is omitted in the plot.}
\end{figure}
A characteristic feature of this dynamical quark model is that the 
critical quark chemical potentials where light and strange quark masses 
jump from their constituent mass values down to almost their current mass 
values do not coincide. With increasing chemical potential the system 
undergoes a sequence of two transitions: (1) vacuum $\to$ two-flavor quark 
matter, (2) two-flavor $\to$ three-flavor quark matter.
The intermediate two-flavor quark matter phase occurs within an interval of
chemical potentials typical for compact star interiors. 
While at intermediate 
coupling the asymmetry between of up and down quark chemical potentials 
leads to a mixed NQ-2SC phase below temperatures of 20-30 MeV, at strong 
coupling the pure 2SC phase extends down to T=0. 
Simultaneously, the limiting chemical potentials of the two-flavor 
quark matter region are lowered by about 40 MeV. 
Three-flavor quark matter 
is always in the CFL phase where all quarks are paired.
The robustness of the 2SC condensate under compact star constraints, with 
respect to changes of the coupling strength, as well as to a softening of the
momentum cutoff by a formfactor, has been recently investigated within a 
different parametrization \cite{Aguilera:2004ag} with similar trend: 
for $\eta=0.75$ and NJL formfactor the 2SC condensate does not occur for 
moderate chemical potentials while for $\eta=1.0$ it occurs simultaneously 
with chiral symmetry  restoration.  
Fig. 3 shows the corresponding dependences of the chemical potentials
conjugate to electric ($\mu_Q$) and color ($\mu_8$) charges.  
All phases considered
in this work have zero $n_3$ color charge for $\mu_3=0$. 
\begin{figure}[htb]
\vspace{-5mm}  
\includegraphics[width=0.45\textwidth,angle=-90]{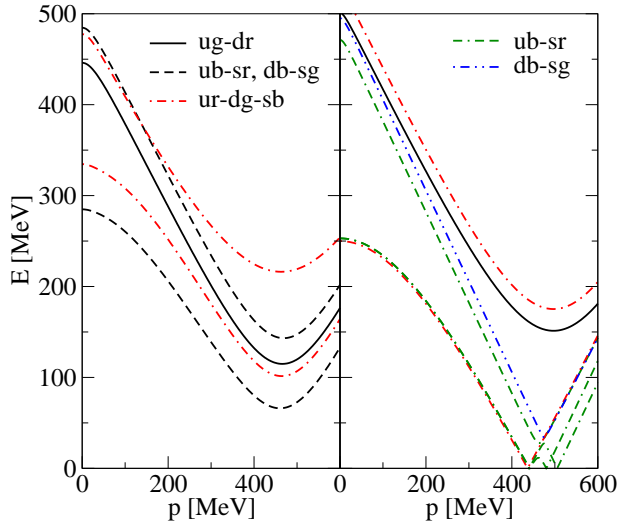}
\caption{Quark-quark quasiparticle dispersion relations.
For $\eta=0.75$, $T=0$, and $\mu=480$~MeV (left panel)
there is a forbidden energy band above the Fermi surface.
All dispersion relations are gapped at this point in the
$\mu-T$ plane, see Fig.~5.
There is no forbidden energy band for the $ub-sr$,
$db-sq$, and $ur-dg-sb$ quasiparticles at $\eta=1$,
$T=84$~MeV, and $\mu=500$~MeV (right panel). This point
in the $\mu-T$ plane constitute a part of the gapless
CFL phase of Fig.~6.
}
\end{figure}
\subsection{Dispersion relations and gapless phases}
In Fig. 4 we show the quasiparticle dispersion relations
of different excitations at two points in the phase diagram:
(I) the CFL phase (left panel), where there is a finite
energy gap for all dispersion relations. (II) the gCFL
phase (right panel), where the energy spectrum is shifted
due to the assymetry in the chemical potentials, such that
the CFL gap is zero and (gapless) excitations with zero
energy are possible. In the present model, this phenomenon
occurs only at rather high temperatures, where the condensates
are diminished by thermal fluctuations.
\begin{figure}[htb]
\vspace{-5mm}
\label{PHASEDIAG_FOR_ETA075}
\includegraphics[width=0.45\textwidth,angle=-90]{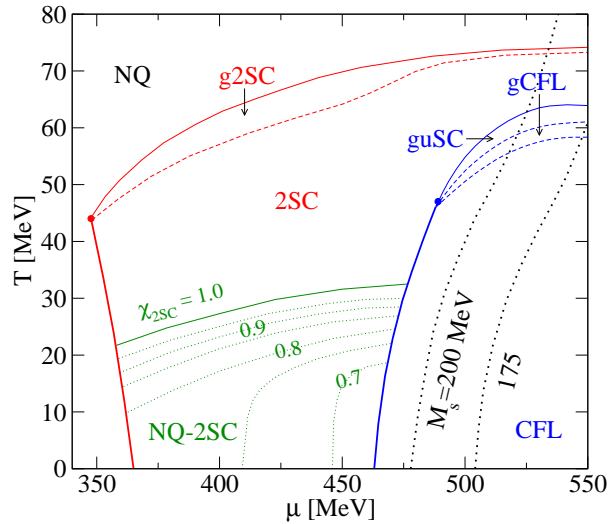}
\caption{Phase diagram of neutral three-flavor quark matter for
intermediate diquark coupling $\eta=0.75$.
First-order phase transition boundaries are indicated by bold solid lines, 
while thin solid lines correspond to second-order phase boundaries. 
The dashed lines indicate gapless phase boundaries.  
The volume fraction, $\chi_{2SC}$, of the 2SC component
of the mixed NQ-2SC phase is denoted with thin dotted
lines, while the constituent strange quark mass is
denoted with bold dotted lines.
 }
\end{figure}
\subsection{Phase diagram}
The thermodynamical state of the system is characterized by the values
of the order parameters and their dependence on T and $\mu$.
Here we illustrate this dependency in a phase diagram. 
We identify the following phases:
\begin{enumerate}
\item NQ: $\Delta_{ud}=\Delta_{us}=\Delta_{ds}=0$;
\item NQ-2SC: $\Delta_{ud}\neq 0$, $\Delta_{us}=\Delta_{ds}=0$, 
0<$\chi_{\rm 2SC}$<1;
\item 2SC: $\Delta_{ud}\neq 0$, $\Delta_{us}=\Delta_{ds}=0$;
\item uSC: $\Delta_{ud}\neq 0$, $\Delta_{us}\neq 0$, $\Delta_{ds}=0$;
\item CFL: $\Delta_{ud}\neq 0$, $\Delta_{ds}\neq 0$, $\Delta_{us}\neq 0$;
\end{enumerate} 
and their gapless versions. 
\begin{figure}[htb]
\vspace{-5mm}
\label{PHASEDIAG_FOR_ETA1}
\includegraphics[width=0.45\textwidth,angle=-90]{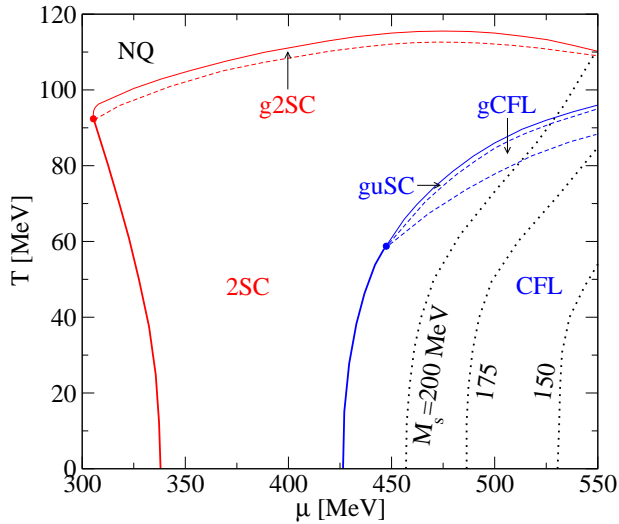}
\caption{Phase diagram of neutral three-flavor quark matter for
strong diquark coupling $\eta=1$. Line styles as in Fig. 3}
\end{figure}
The resulting phase diagrams for intermediate 
and strong coupling are given in Figs. 5 and 6, resp. and constitute the main 
result of this work, which is summarized in the following statements:
\begin{enumerate}
\item Gapless phases occur only at high temperatures, 
above 50 MeV (intermediate coupling) or 60 MeV (strong coupling).
\item CFL phases occur only at rather high chemical potential, well above 
the chiral restoration transition, i.e. above 464 MeV
 (intermediate coupling) or 426 MeV (strong coupling).
\item Two-flavor quark matter for intermediate coupling is at low temperatures 
(T<20-30 MeV) in a mixed NQ-2SC phase, at high temperatures in the pure 2SC 
phase.
\item Two-flavor quark matter for strong coupling is in the 2SC phase with
rather high critical  temperatures of $\sim 100$ MeV.
\item The critical endpoint of first order chiral phase transitions is at 
(T,$\mu$)=(44 MeV, 347 MeV) for intermediate coupling and at (92 MeV, 305 MeV)
for strong coupling.
\end{enumerate}

\subsection{Quark matter equation of state}
The various phases of quark matter presented in the previous
section have been identified by minimizing the thermodynamic
potential, $\Omega$, in the order parameters, $\Delta_{ij}$
and $\phi_i$. For a homogenous system, the pressure is
$P=-\Omega_{min}$, see Fig. 7, where the $\mu$-dependence
of $\Omega_{min}$ is shown at $T=0$ for the different
competing phases. The lowest value of $\Omega_{min}$
corresponds to the negative value of the physical pressure.
The intersection of two curves corresponds to a first
order phase transition. All other thermodynamic quantities
can be obtained from the thermodynamic potential by
derivatives. At intermediate coupling, we have a first order
transition from the NQ-2SC phase to the CFL phase, whereas
at strong coupling the first order transition is from
the 2SC phase to the CFL phase, with a lower critical
energy density. In Fig. 8 the equation of state for cold
three-flavor quark matter is given on a form suitable for
the investigation of hydrodynamic stability of gravitating
compact object, so-called quark stars. This is the topic
for the next Subsection.
\begin{figure}[th]
\vspace{-5mm}
\includegraphics[width=0.45\textwidth,angle=-90]{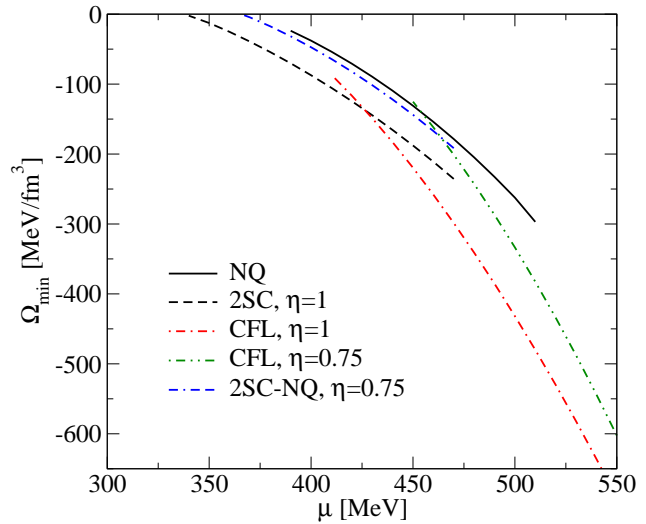}
\caption{Minima of the thermodynamical potential for neutral three-flavor 
quark matter at T=0 as a function of the quark chemical potential.
Note that at a given coupling $\eta$ the state with the lowest 
$\Omega_{\rm min}$ 
is attained and the physical pressure is $P=-\Omega_{\rm min}$.}
\end{figure}
\begin{figure}[h]
\vspace{-3mm}
\includegraphics[width=0.45\textwidth,angle=-90]{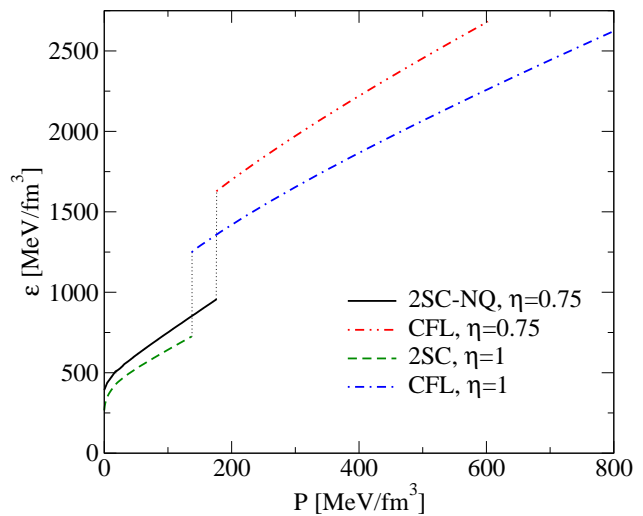}
\vspace{-1mm}
\caption{Equation of state for three-flavor quark matter at T=0 with first
order phase transitions. For intermediate diquark coupling ($\eta=0.75$):
from the mixed NQ-2SC phase to the CFL phase, 
for strong diquark coupling ($\eta=1$):
from the 2SC phase to the CFL phase.}
\end{figure}
\subsection{Quark star configurations}
The properties of spherically symmetric, static
configurations of dense matter can be calculated
with the well-known Tolman-Oppenheimer-Volkoff
equations for hydrostatic equilibrium of self-
gravitating matter, see also \cite{glendenning00},
\begin{equation}
\frac{dP(r)}{dr}= -\frac{[\varepsilon(r)+P(r)][m(r)+4\pi
r^{3}P(r)]}{r[r-2m(r)]}~.
\end{equation}

Here $\varepsilon(r)$ is the energy density and $P(r)$ the pressure at 
distance $r$ from the center of the star. The mass enclosed in a sphere with
radius $r$ is defined by
\begin{equation}
m(r)=4\pi \int_{0}^{r}\varepsilon(r')r'^{2}dr'~.
\end{equation}

These equations are solved for given central baryon
number densities, $n_B(r=0)$, thereby defining a
sequence of quark star configurations.
For the generalization to finite temperature configurations, see 
\cite{ST}. Hot quark stars have been discussed, e.g., in
\cite{Kettner:1994zs,Blaschke:1998hy,Blaschke:2003yn}.
\begin{figure}[ht]
\label{stab}
\vspace{-3mm}
\includegraphics[width=0.45\textwidth,angle=-90]{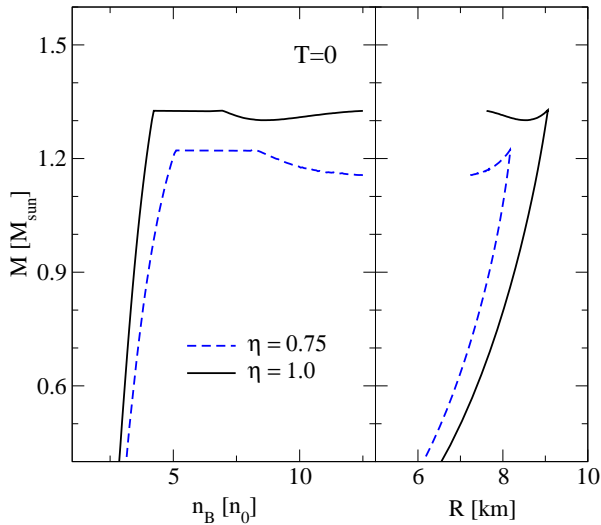}
\vspace{-2mm}
\caption{Sequences of cold quark stars for the three-flavor quark matter 
equation of state described in the text. 
The rising branches in the mass-central density relation (left panel)
indicate stable compact object configurations. 
The mass-radius relations (right panel) show that the three-flavor 
quark matter described in this paper leads to very compact selfbound objects. 
For intermediate diquark coupling, $\eta=0.75$, stable stars 
consist of a mixed phase of NQ-2SC matter with a maximum mass of 
$1.21$ M$_\odot$ (dashed line).
At higher densities a phase transition to CFL quark matter occurs which 
entails a collapse of the star.
For strong coupling, $\eta=1$, the low density quark matter is in 
the 2SC phase and corresponding quark stars are stable up to a 
maximum mass of 
$1.326~M_\odot$ (solid line).
The phase transition to CFL quark matter entails an instability which 
at T=0 leads to a third family of stable stars for central densities above 
9 n$_0$ and a mass twin window of  1.301 - 1.326 $M_\odot$.}
\end{figure} 
In Fig. 9 we show the stable configurations of
quark stars for the three-flavor quark matter
equation of state described above. The obtained
mass radius relations allow for very compact
selfbound objects, with a maximum radius that
is less than 10~km. For intermediate diquark
coupling, $\eta=0.75$, stable stars consist of
a NQ-2SC mixed phase with a maximum mass of
$1.21~M_\odot$. With incrasing density, a phase
transition to the CFL phase renders the sequence 
unstable. For the strong diquark coupling,
$\eta=1$, quark matter is in the 2SC phase at
low densities and the corresponding sequence
of quark stars is stable up to a maximum
mass of 1.33~$M_\odot$. The phase transition
to CFL quark matter entails an instability
that leads to a third family of stable stars,
with masses in-between 1.30 and $1.33~M_\odot$.
\begin{figure}[ht]
\vspace{-5mm}
\label{mdef1}
\includegraphics[width=0.45\textwidth,angle=-90]{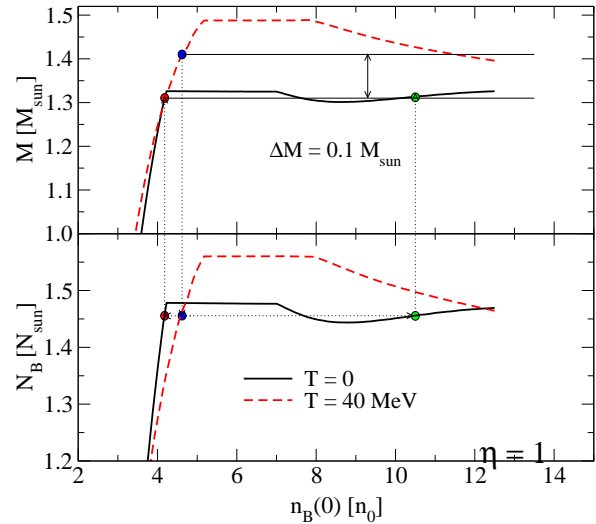}
\caption{Cooling an isothermal quark star configuration with initial 
mass $M=1.41~M_\odot$ at temperature $T=40$ MeV under conservation of
the given baryon number $N=1.46~N_\odot$ down to $T=0$ leads to a mass defect 
$\Delta M=0.1~M_\odot$ for the strong coupling case ($\eta=1.0$).
Due to the twin structure at $T=0$, two alternatives for the final state can
be attained, a homogeneous 2SC quark star or a dense 2SC-CFL quark hybrid star.
 } 
\end{figure}
\begin{figure}[ht]
\vspace{-5mm}
\label{struc}
\includegraphics[width=0.45\textwidth,angle=-90]{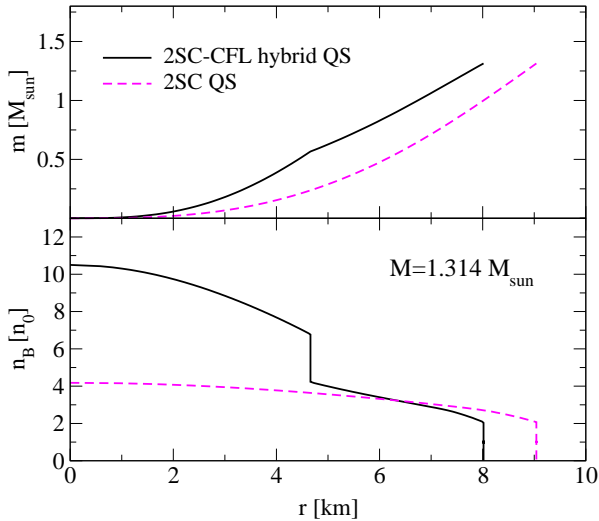}
\caption{Structure of two quark star (QS) configurations with 
$M=1.314~M_\odot$ (mass twins) for the three-flavor quark matter 
equation of state described in the text in the case of strong coupling 
($\eta=1$). 
The  low-density twin has a radius of 9 km and is a homogeneous 2SC quark star,
the high-density twin is more compact with a radius of 8 km and consists of 
a CFL quark matter core with 4.65 km radius and a 2SC quark matter shell.}
\end{figure}
For non-accreting compact stars the baryon number
is an invariant during the cooling evolution. By
comparing the masses of cold and hot isothermal
configurations of quark stars of equal baryon
number, the maximum mass defect (energy release
due to cooling) can be calculated. The result for
the strong diquark coupling, $\eta=1$, is shown
if Fig. 10. For an initial temperature of 40~MeV
and a given baryon number of $N=1.46~N_\odot$, the
initial mass is $M=1.41~M_\odot$.
By cooling this object down to $T=0$, a mass defect of 
$\Delta M = 0.1~M_\odot$ occurs. 
For the chosen baryon number, $N=1.46~N_\odot$,
there are two possible $T=0$ configurations (twins).
A hot star could thus evolve into the more compact
mass-equivalent (twin) final state, if a fluctuation
triggers the transition to a CFL phase in the core
of the star. The structures of these two twin
configurations are given in Fig. 11. The energy
release of 0.1~$M_\odot$ is of the same order of
magnitude as the energy release in supernova
explosions and gamma-ray bursts. Disregarding
the possible influence of a hadronic shell and
the details regarding the heat transport, the
cooling induced first order phase transition
to the CFL phase 
could serve as a candidate process for the
puzzling engine of these energetic phenomena 
\cite{Aguilera:2004ag,Blaschke:2003yn}.


\section{Conclusions}
We have investigated the phase diagram of three-flavor
quark matter within an NJL model under compact star
constraints. Local color and electric charge neutrality
is imposed for $\beta-$equilibrated superconducting
quark matter. The constituent quark masses are
selfconsistently determined. The model refrains from
adopting the 't Hooft determinant interaction in the
mean field Lagrangian as a realization of the U$_A$(1)
symmetry breaking.
Instead, it is assumed that the $\eta-\eta'$ mass difference
originates from an anomalous coupling of the pseudoscalar
isosinglet fluctuation to the nonperturbative gluon sector,
which gives no contribution to the quark thermodynamics at
the mean field level. The resulting parametrization of this
SU$_f$(3) NJL model results in a stronger coupling than NJL
models with a 't Hooft term and thus in different phase
diagrams, cf. Ref. \cite{RWBSR}. The diquark condensates are
determined selfconsistently by minimization of the grand canonical
thermodynamic potential.
The various condensates are order parameters that
characterize the different phases in the plane of temperature
and quark chemical potential. These phases are in particular
the NQ-2SC mixed phase, the 2SC, uSC, and CFL phases, as well
as the corresponding gapless phases.
We have investigated strong and intermediate diquark coupling
strengths. It is shown that in both cases gapless superconducting
phases does not occur at temperatures relevant for compact star
evolution, i.e., below $\sim 50$~MeV. Three-flavor quark matter
phases, e.g., the CFL phase, occur only at rather large
chemical potential, so the existence of such phases in stable
compact stars is questionable. The stability and stucture of
isothermal quark star configurations are evaluated. For the
strong diquark coupling, 2SC stars are stable up to a maximum
mass of $1.33~M_\odot$. A second family of more compact
stars (twins) with a CFL quark matter core and masses in-between
$1.30$ and $1.33~M_\odot$ are found to be stable. For intermediate
coupling, the quark stars are composed of a mixed NQ-2SC phase up
to a maximum mass of $1.21~M_\odot$, where a phase transition
to the CFL phase occurs and the configurations become unstable.
When isothermal star configurations with an initial temperature
of 40~MeV cools under conservation of baryon number, the mass defect
is $0.1~M_\odot$ for the strong diquark coupling.
It is important to investigate the robustness of these statements,
in particular by inlcuding nonlocal formfactors and by going beyond
the mean-field level by including the effects of a hadronic
medium on the quark condensates. Finally, any statement concerning
the occurence and stability of quark matter in compact stars shall
include an investigation of the influence of a hadronic shell
\cite{Baldo:2002ju,Shovkovy:2003ce,Grigorian:2003vi}
on the solutions of the equations of compact star structure.


\section{Acknowledgements}

F.S. acknowledges support from the Swedish National Graduate School of 
Space Technology. A.M.\"O. received support from Hacettepe University 
Research Fund, grant No. 02 02 602 001. D.B. thanks for partial support 
of the Department of Energy during the program INT-04-1 on {\it QCD and 
Dense Matter: From Lattices to Stars} at the University of Washington, 
where this work has been started.
This work has been supported in part by the Virtual Institute of the Helmholtz 
Association under grant No. VH-VI-041.
We are grateful to our colleagues in Darmstadt and Frankfurt who made the 
results of their study in Ref. \cite{RWBSR} available to us prior to 
submission.


\appendix
\section{Dispersion Relations}
The dispersion relations of the quasiparticles that appear in the expression 
for the thermodynamic potential~\eqn{potential} is the eigenvalues of the
Nambu-Gorkov matrix~\eqn{eigmatrix}. For each color and flavor combination of
the eight component Nambu-Gorkov spinors, there is a corresponding 8x8 entry in
this matrix; for three flavors and three colors~\eqn{eigmatrix} is a 
72x72 matrix.
The explicit form of this matrix can be represented by a table, where the rows
and columns denote the flavor and color degrees of freedom
\begin{widetext}
\begin{equation}
\label{matrix72}
      \setlength\arraycolsep{-0.03cm}
     \begin{array}{c|cccccccccccccccccc}
        &q_{ur}& q_{ug}&q_{ub}&q_{dr}& q_{dg}&q_{db}&q_{sr}& q_{sg}&q_{sb}&q_{ur}^\dagger& q_{ug}^\dagger&q_{ub}^\dagger&q_{dr}^\dagger& q_{dg}^\dagger&q_{db}^\dagger&q_{sr}^\dagger& q_{sg}^\dagger&q_{sb}^\dagger \\
        \hline
        q_{ur}^\dagger\; &\;\;A_{ur}&0&0&0&0&0&0&0&0&0&0&0&0&D_{ud}&0&0&0&D_{us} \\
        q_{ug}^\dagger\; &0&A_{ug}&0&0&0&0&0&0&0&0&0&0&-D_{ud}&0&0&0&0&0 \\
        q_{ub}^\dagger\; &0&0&A_{ub}&0&0&0&0&0&0&0&0&0&0&0&0&-D_{us}&0&0 \\
        q_{dr}^\dagger\; &0&0&0&A_{dr}&0&0&0&0&0&0&-D_{ud}&0&0&0&0&0&0&0 \\
        q_{dg}^\dagger\; &0&0&0&0&A_{dg}&0&0&0&0&D_{ud}&0&0&0&0&0&0&0&D_{ds} \\
        q_{db}^\dagger\; &0&0&0&0&0&A_{db}&0&0&0&0&0&0&0&0&0&0&-D_{ds}&0 \\
        q_{sr}^\dagger\; &0&0&0&0&0&0&A_{sr}&0&0&0&0&-D_{us}&0&0&0&0&0&0 \\
        q_{sg}^\dagger\; &0&0&0&0&0&0&0&A_{sg}&0&0&0&0&0&0&-D_{ds}&0&0&0 \\
        q_{sb}^\dagger\; &0&0&0&0&0&0&0&0&A_{sb}&\,D_{us}&0&0&0&D_{ds}&0&0&0&0\\
        q_{ur}\;         &0&0&0&0&D^\dagger_{ud}&0&0&0&D^\dagger_{us}\,&B_{ur}&0&0&0&0&0&0&0&0 \\
        q_{ug}\;         &0&0&0&-D^\dagger_{ud}&0&0&0&0&0&0&B_{ug}&0&0&0&0&0&0&0 \\
        q_{ub}\;         &0&0&0&0&0&0&-D^\dagger_{us}&0&0&0&0&B_{ub}&0&0&0&0&0&0 \\
        q_{dr}\;         &0&-D^\dagger_{ud}&0&0&0&0&0&0&0&0&0&0&B_{dr}&0&0&0&0&0 \\
        q_{dg}\;         &\;\;D^\dagger_{ud}&0&0&0&0&0&0&0&D^\dagger_{ds}&0&0&0&0&B_{dg}&0&0&0&0 \\
        q_{db}\;         &0&0&0&0&0&0&0&-D^\dagger_{ds}&0&0&0&0&0&0&B_{db}&0&0&0 \\
        q_{sr}\;         &0&0&-D^\dagger_{us}&0&0&0&0&0&0&0&0&0&0&0&0&B_{sr}&0&0 \\
        q_{sg}\;         &0&0&0&0&0&-D^\dagger_{ds}&0&0&0&0&0&0&0&0&0&0&B_{sg}&0 \\
        q_{sb}\;         &\;\;D^\dagger_{us}&0&0&0&D^\dagger_{ds}&0&0&0&0&0&0&0&0&0&0&0&0&B_{sb}
        \end{array}.
\end{equation}
\end{widetext}
Each entry is a 4x4 Hermitean matrix in Dirac space. 
The diagonal submatrices are
\begin{equation}
\label{Ablock}
        A_{i,\alpha} = \left[
        \setlength\arraycolsep{-0.01cm}
        \begin {array}{cccc}
        p+\mu_{i,\alpha}&0&-M_{{i}}&0 \\
        0&-p+\mu_{i,\alpha}&0&-M_{{i}} \\
        -M_{{i}}&0&-p+\mu_{i,\alpha}&0 \\
        0&-M_{{i}}&0&p+\mu_{i,\alpha}
        \end {array}
        \right], 
\end{equation}
\begin{equation}
\label{Bblock}
        B_{j,\beta} = \left[
        \setlength\arraycolsep{-0.01cm}
        \begin {array}{cccc}
        -p-\mu_{j,\beta}&0&M_{{j}}&0 \\
        0&p-\mu_{j,\beta}&0&M_{{j}}\\
        M_{{j}}&0&p-\mu_{j,\beta}&0\\
        0&M_{{j}}&0&-p-\mu_{j,\beta}
        \end{array}
        \right], 
\end{equation}
whereas the off-diagonal blocks are given by
\begin{equation}
\label{Dblock}
        D_{i,j} = \left[
        \setlength\arraycolsep{-0.01cm}
        \begin {array}{cccc}
        0&0&0&i\Delta_{i,j}\\
        0&0&-i\Delta_{i,j}&0\\
        0&i\Delta_{i,j}&0&0\\
        -i\Delta_{i,j}&0&0&0
        \end{array}
        \right].
\end{equation}

The eigenvalues of~\eqn{matrix72} are the quasiparticle energies, $\lambda_a$, that
enter the thermodynamic potential~\eqn{potential}, {\it i.e.}, the 72
dispersion relations of the various quark-quark and antiquark-antiquark excitations.
These eigenvalues can be calculated using a standard numerical library. 
However, in order to reduce the computational cost, the matrix can be 
decomposed into a block-diagonal matrix by elementary row and column 
operations.
\begin{widetext}
\begin{equation}
        \label{blockmatrix72}
        \setlength\arraycolsep{-0.03cm}
        \begin{array}{c|cccccccccccccccccc}
                &q_{ur}&q_{dg}&q_{sb}&q^\dagger_{ur}&q^\dagger_{dg}&q^\dagger_{sb}
                &q_{ug}&q_{dr}^\dagger&q_{dr}&q_{ug}^\dagger&q_{ub}&q_{sr}^\dagger
                &q_{sr} &q_{ub}^\dagger&q_{db}&q_{sg}^\dagger&q_{sg}&q_{db}^\dagger \\
                \hline
                q_{ur}^\dagger\;&\;\;A_{ur}&0&0&0&D_{ud}&\;D_{us}&0&0&0&0&0&0&0&0&0&0&0&0 \\
                q_{dg}^\dagger\;&0&A_{dg}&0&D_{ud}&0&D_{ds}&0&0&0&0&0&0&0&0&0&0&0&0 \\
                q_{sb}^\dagger\;&0&0&A_{sb}&D_{us}&\;D_{ds}&0&0&0&0&0&0&0&0&0&0&0&0&0 \\
                q_{ur}&0&D^\dagger_{ud}&\;D^\dagger_{us}&\;B_{ur}&0&0&0&0&0&0&0&0&0&0&0&0&0&0 \\
                q_{dg}&D^\dagger_{ud}&0&D^\dagger_{ds}&0&B_{dg}&0&0&0&0&0&0&0&0&0&0&0&0&0 \\
                q_{sb}&\;D^\dagger_{us}&\;D^\dagger_{ds}&0&0&0&B_{sb}&0&0&0&0&0&0&0&0&0&0&0&0 \\
                q_{ug}^\dagger\;&0&0&0&0&0&0&A_{ug}&-D_{ud}&0&0&0&0&0&0&0&0&0&0 \\
                q_{dr}&0&0&0&0&0&0&-D^\dagger_{ud}&B_{dr}&0&0&0&0&0&0&0&0&0&0 \\
                q_{dr}^\dagger\;&0&0&0&0&0&0&0&0&A_{dr}&-D_{ud}&0&0&0&0&0&0&0&0 \\
                q_{ug}&0&0&0&0&0&0&0&0&-D^\dagger_{ud}&B_{ug}&0&0&0&0&0&0&0&0 \\
                q_{ub}^\dagger\;&0&0&0&0&0&0&0&0&0&0&A_{ub}&-D_{us}&0&0&0&0&0&0 \\
                q_{sr}&0&0&0&0&0&0&0&0&0&0&-D^\dagger_{us}&B_{sr}&0&0&0&0&0&0 \\
                q_{sr}^\dagger\;&0&0&0&0&0&0&0&0&0&0&0&0&A_{sr}&-D_{us}&0&0&0&0 \\
                q_{ub}&0&0&0&0&0&0&0&0&0&0&0&0&-D^\dagger_{us}&B_{ub}&0&0&0&0 \\
                q_{db}^\dagger\;&0&0&0&0&0&0&0&0&0&0&0&0&0&0&A_{db}&-D_{ds}&0&0 \\
                q_{sg}&0&0&0&0&0&0&0&0&0&0&0&0&0&0&-D^\dagger_{ds}&B_{sg}&0&0 \\
                q_{sg}^\dagger\;&0&0&0&0&0&0&0&0&0&0&0&0&0&0&0&0&A_{sg}&-D_{ds} \\
                q_{db}&0&0&0&0&0&0&0&0&0&0&0&0&0&0&0&0&-D^\dagger_{ds}&B_{db} \\
\end{array}
\end{equation}
\end{widetext}

This matrix has one 24x24 and six 8x8 independent submatrices. Expressing these
submatrices explicitely, using~\eqns{Ablock}{Dblock}, the 24x24 matrix can
be decomposed into two independent 12x12 submatrices by elementary row and
column operations. Similarly, the six 8x8 matrices can be transformed into
twelve independent 4x4 submatrices. There is a two-fold degeneracy due to the
Nambu-Gorkov basis, each matrix appears both as ${\cal M}$ and ${\cal M}^\dagger$,
so there are only one independent 12x12 matrix and six 4x4 matrices. The 12x12 matrix is
\begin{widetext}
\begin{equation}
        \label{matrix12}
        \setlength\arraycolsep{-0.12cm}
        {\cal M}_{12} = \left[ \begin {array}{cccccccccccc}
        \; p+\mu_{ur}&0&0&-M_{u}&0&0&0&i\Delta_{ud}&i\Delta_{us}&0&0&0\\
        0&p+\mu_{dg}&0&0&-M_{d}&0&i\Delta_{ud}&0&i\Delta_{ds}&0&0&0\\
        0&0&p+\mu_{sb}&0&0&-M_{s}&i\Delta_{us}&i\Delta_{ds}&0&0&0&0\\
        -M_{u}&0&0&-p+\mu_{ur}&0&0&0&0&0&0&i\Delta_{ud}&i\Delta_{us}\\
        0&-M_{d}&0&0&-p+\mu_{dg}&0&0&0&0&i\Delta_{ud}&0&i\Delta_{ds}\\
        0&0&-M_{s}&0&0&-p+\mu_{sb}&0&0&0&i\Delta_{us}&i\Delta_{ds}&0\\
        0&\;\;-i\Delta_{ud}&\;\;\;-i\Delta_{us}&0&0&0&-p-\mu_{ur}&0&0&M_{u}&0&0\\
        -i\Delta_{ud}&0&-i\Delta_{ds}&0&0&0&0&-p-\mu_{dg}&0&0&M_{d}&0\\
        -i\Delta_{us}&-i\Delta_{ds}&0&0&0&0&0&0&-p-\mu_{sb}&0&0&M_{s}\\
        0&0&0&0&-i\Delta_{ud}&-i\Delta_{us}&M_{u}&0&0&p-\mu_{ur}&0&0\\
        0&0&0&-i\Delta_{ud}&0&-i\Delta_{ds}&0&M_{d}&0&0&p-\mu_{dg}&0\\
        0&0&0&-i\Delta_{us}&-i\Delta_{ds}&0&0&0&M_{s}&0&0&p-\mu_{sb}\;
        \end{array}\right],
\end{equation}
\end{widetext}
and the 4x4 matrices are
\begin{equation}
        \label{matrix4}
        {\cal M}_4 = \left[
        \setlength\arraycolsep{-0.01cm}
        \begin {array}{cccc}
        p+\mu_{i,\alpha}&-i\Delta_{i,j}&-M_{i}&0\\
        i\Delta_{i,j}&-p-\mu_{j,\beta}&0&M_{j}\\
        -M_{i}&0&-p+\mu_{i,\alpha}&-i\Delta_{i,j}\\
        0&M_{j}&i\Delta_{i,j}&p-\mu_{j,\beta}
        \end {array}
        \right],
\end{equation}
for spinor products $ug-dr$, $ub-sr$, $db-sg$, $dr-ug$, $sr-ub$, and $sg-db$,
respectively.

Since these matrices are Hermitean, the eigenvalues appear
in $\pm$ pairs. Thus, in general, there are nine independent dispersion
relations for quark-quark excitations and nine for antiquark-antiquark
excitations. The eigenvalues of~\eqn{matrix12} must be calculated
numerically. The eigenvalues of~\eqn{matrix4} can be obtained
analytically by solving for the roots of the quartic characteristic
polynomial,
\begin{equation}
\label{quartic}
        \lambda^4 + a_3 \lambda^3 + a_2 \lambda^2 + a_1 \lambda + a_0=0,
\end{equation}
where
\begin{eqnarray}
        a_0 &=& P^4+\left(M^2_i+M^2_j+2\Delta^2_{i,j}-\mu^2_{i,\alpha}-\mu^2_{j,\beta}\right)P^2 \nonumber \\
        &+&\left(\mu_{i,\alpha}\mu_{j,\beta}+M_iM_j+\Delta^2_{i,j}+\mu_{i,\alpha}M_j+\mu_{j,\beta}M_i\right) \nonumber \\
        &&\left(\mu_{i,\alpha}\mu_{j,\beta}+M_iM_j+\Delta^2_{i,j}-\mu_{i,\alpha}M_j-\mu_{j,\beta}M_i\right), \nonumber \\
        a_1 &=& 2\left(\mu_{i,\alpha}-\mu_{j,\beta}\right)P^2
        +2\Delta^2_{i,j}\left(\mu_{i,\alpha}-\mu_{j,\beta}\right) \nonumber \\
        &+&\,2\left(\mu_{i,\alpha}M^2_j-\mu_{j,\beta}M^2_i
        +\mu^2_{i,\alpha}\mu_{j,\beta}-\mu^2_{j,\beta}\mu_{i,\alpha}\right), \nonumber \\
        a_2 &=& \mu^2_{i,\alpha}+\mu^2_{j,\beta}-2P^2-M^2_i-M^2_j-2\Delta^2_{i,j}
        -4\mu_{i,\alpha}\mu_{j,\beta}, \nonumber \\
        a_3 &=& -2\left(\mu_{i,\alpha}-\mu_{j,\beta}\right). \nonumber
\end{eqnarray}

In the limit when $M_i=M_j=M$, which is approximately valid for the $ug-dr$ and $dr-ug$
quasiparticles, the four solutions are
\begin{equation}
        \lambda = \frac{\mu_{i,\alpha}-\mu_{j,\beta}}{2}
                \pm \sqrt{\left(\frac{\mu_{i,\alpha}+\mu_{j,\beta}}{2} \pm E\right)^2+\Delta^2_{i,j}},
\end{equation}
where $E=\sqrt{p^2+M^2}$. This result is in agreement with~\cite{RWBSR}. More generally,
the solutions of the quartic equation can be found in textbooks, see,
{ e.g.},~\cite{Abramowitz:1972}. In this work the eigenvalues of the 4x4 matrices
were calculated with the exact solutions of the quartic equation and the eigenvalues of
the 12x12 matrix were calculated with LAPACK. The momentum integral in~\eqn{potential}
was calculated with a Gaussian quadrature. The minimization of the thermodynamic
potential was performed with conjugate gradient methods, choosing the initial values
of the variational parameters carefully, and then comparing the free energies of the
various minima. The color and electric charges were neutralized with a globally
convergent Newton-Raphson method in multidimensions.

Gapless excitations/dispersion relations are characterized by a non-zero condensate,
$\Delta_{i,j}$, and a corresponding dispersion relation that is zero for at least one
value of the quasiparticle momentum, {\it i.e.}, the dispersion relation reaches the
Fermi surface and there is no forbidden energy band.



\end{document}